\documentclass[%
 aip,rsi,amsmath,amssymb,reprint,longbibliography]{revtex4-1}

\usepackage{graphicx}
\usepackage{dcolumn}
\usepackage{bm}
\usepackage{gensymb}
\usepackage{color}

\begin{document}

\preprint{AIP/123-QED}

\title{Injection locking and parametric locking in a superconducting circuit}

\author{D. Markovi\'c}
\affiliation{Laboratoire Pierre Aigrain, D\'epartement de physique de l'ENS, \'Ecole normale sup\'erieure, PSL Research University, Univ. Paris Diderot, Sorbonne Paris Cit\'e, Sorbonne Universit\'es, UPMC Univ. Paris 06, CNRS, 75005 Paris, France}
\author{J.D. Pillet}
\affiliation{Laboratoire des Solides Irradi\'es, \'Ecole Polytechnique, CNRS, CEA, 91128 Palaiseau, France}
\author{E. Flurin}
\affiliation{SPEC, CEA, CNRS, Univ. Paris-Saclay, CEA Saclay, 91191 Gif-sur-Yvette, France}
\author{N. Roch}
\affiliation{Univ. Grenoble Alpes, CNRS, Grenoble INP, Institut N\'eel,
25 rue des Martyrs BP 166, 38042 Grenoble, France}
\author{B. Huard}
\affiliation{Univ. Lyon, ENS de Lyon, Univ. Claude Bernard, CNRS, Laboratoire de Physique, F-69342 Lyon, France}
\affiliation{Laboratoire Pierre Aigrain, D\'epartement de physique de l'ENS, \'Ecole normale sup\'erieure, PSL Research University, Univ. Paris Diderot, Sorbonne Paris Cit\'e, Sorbonne Universit\'es, UPMC Univ. Paris 06, CNRS, 75005 Paris, France}

\begin{abstract}
When a signal is injected in a parametric oscillator close enough to its resonance, the oscillator frequency and phase get locked to those of the injected signal. Here, we demonstrate two frequency locking schemes using a Josephson mixer in the parametric down-conversion regime, pumped beyond the parametric oscillation threshold. The circuit then emits radiation out of two spectraly and spatially separated resonators at frequencies determined by the locking schemes that we choose. When we inject the signal close to a resonance, it locks the oscillator emission to the signal frequency by injection locking. When we inject the signal close to the difference of resonances, it locks the oscillator emission by parametric locking. We compare both schemes and investigate the dependence of the parametric locking range on the pump and the injection signal power. Our results can be interpreted using Adler's theory for lasers, which makes a new link between laser physics and superconducting circuits that could enable better understanding of pumped circuits for quantum information applications such as error correction, circulators and photon number detectors.
\end{abstract}

\maketitle

\section{Introduction}

Injection locking is a phenomenon through which emission frequency and phase of a parametric oscillator become locked on those of an injection tone. It is usually performed by injection of a weak AC signal close to the natural oscillator frequency and has given rise to applications including narrowing the linewidth of lasers~\cite{Stover1966, Hillbrand2019}, driving CMOS
based oscillators~\cite{Razavi2004}, understanding synchronization of biological systems~\cite{Mirolli1990} and neuromorphic computing~\cite{Romera2018, Markovic2019}. Injection locking has recently been experimentally realized in various mesoscopic devices~\cite{Rippard2005,Liu2015}, including micromechanical oscillators~\cite{Zhang2012} and superconducting circuits~\cite{Cassidy2017,Bengtsson2018a}.

In this letter, we demonstrate parametric locking of two non-degenerate oscillators with resonance frequencies $\omega_a$ and $\omega_b$, by a signal injected close to the difference of their frequencies $\omega_a -\omega_b$. The oscillators are coupled by a Josephson circuit pumped at $\omega_p = \omega_a + \omega_b$. We show that parametric locking relies on multiphotonic processes taking place in the Josephson circuit and we compare it to the injection locking by a signal injected close to the natural frequency $\omega_a$ or $\omega_b$ of one of the oscillators. We illustrate the differences between these two approaches by measuring the frequency and power dependence of the emission spectra of the two resonators and describe our observations using an extension of Adler's theory \cite{Adler1973}.


\section{Measurement setup and device parameters}

\begin{figure}
\includegraphics[scale=0.45]{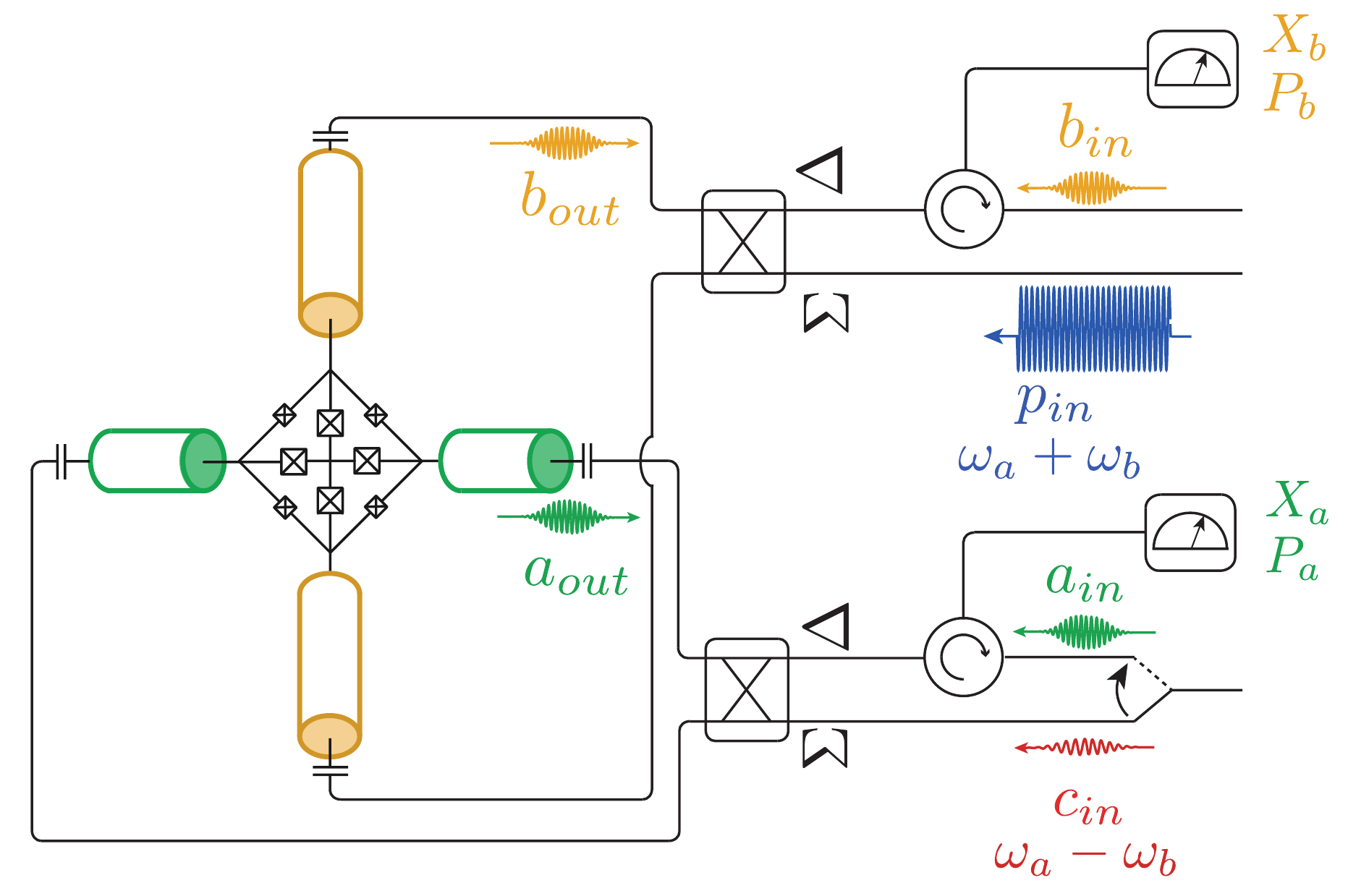}
\caption{The Josephson mixer consists of two microwave resonators $a$ (orange) and $b$ (green) resonating at different frequencies $\omega_a$ and $\omega_b$ that are coupled via a Josephson Ring Modulator. Two 180$\degree$  hybrid couplers (boxed crossings symbols) are used to inject microwave signals in both differential and common driving on $a$ and $b$ ports. The circuit can be  pumped at amplitude $p_{in}$ and frequency $\omega_p = \omega_a + \omega_b$  above the parametric oscillation threshold. An additional locking drive is injected either on mode $a_{in}$ at frequency $\omega_{in,a}\approx\omega_a$, or on mode $b_{in}$ at frequency $\omega_{in,b}\approx\omega_b$, or on the common mode $c_{in}$ at frequency $\omega_c \approx \omega_a-\omega_b$.}
\label{schema}
\end{figure}
 In our devices, the resonators $a$ and $b$ are the $\lambda/2$ modes of two aluminum superconducting microstrip lines arranged in a cross shape. They resonate at frequencies $\omega_a=2\pi\times 8.445~\mathrm{GHz}$ and $\omega_b=2\pi\times 6.451~\mathrm{GHz}$. They are physically connected in their center by a Josephson ring modulator (JRM, Fig.~\ref{schema})~\cite{Bergeal2010,Roch2012, Abdo2013a}, which couples them when their common mode is pumped far from resonance with an external signal of amplitude $p$. The Hamiltonian of the circuit is then given by
 \begin{equation}
\hat{H} = \hbar \omega_a \hat{a}^\dagger \hat{a} + \hbar \omega_b \hat{b}^\dagger \hat{b} + \hat{H}_{\textrm{JRM}},\label{H3WM}
 \end{equation}
where, at the lowest order in the fields, the Josephson ring modulator Hamiltonian reads~\cite{Bergeal2010, Abdo2013a}
\begin{equation}
\hat{H}_{\textrm{JRM}} = \hbar \chi (p +p^\ast)(\hat{a}+\hat{a}^\dagger)(\hat{b}+\hat{b}^\dagger).\label{HJRM}
\end{equation}

When the pump is applied at frequency $\omega_p = \omega_a + \omega_b$, this three-wave mixing Hamiltonian simplifies to the parametric down-conversion Hamiltonian~\cite{Abdo2013a, Markovic2018} in the rotating wave approximation as $\hat{H}_\textrm{pdc}=\hbar \chi(p\hat{a}^\dagger \hat{b}^\dagger + p^\ast \hat{a}\hat{b})$. Two regimes can be distinguished as a function of pump power. Below the so-called parametric oscillation threshold $(p_{in}<p_{th})$, the circuit behaves as a non-degenerate amplifier. Without any input drives on $a$ and $b$ ports, the circuit amplifies vacuum fluctuations into the output modes $a_{out}$ and $b_{out}$. More precisely, it generates a vacuum two-mode squeezed state non-degenerate in frequency and space~\cite{Flurin2012}, whose squeezing ratio increases with the pump power. The threshold power corresponds to a conversion rate of pump photons into $a$ and $b$ photons that is as large as the geometric mean of the dissipation rates $\kappa_a$ and $\kappa_b$ of the modes $a$ and $b$, such that the cooperativity $C=4 \chi^2|p_{th}|^2/ (\kappa_a \kappa_b)$ is equal to 1. Beyond this threshold, the device enters the parametric oscillation regime characterized by a spontaneous generation of photons in the two resonators and correlated emissions~\cite{Kamal2009} out of modes $a$ and $b$. In absence of mechanisms limiting the resonators population, this regime is unstable as the number of photons keeps increasing. There exist typically two processes which can stabilize parametric oscillations. First, the pump power can be depleted as down-conversion becomes stronger, such that
the pump cannot be considered stiff anymore and $p$ does not increase with $p_{in}$ any longer. It is this process that usually stabilizes lasing in optics~\cite{Reynaud87, Milburn}. Second, Kerr nonlinearities can shift the mode frequencies when the pump power increases and they ultimately limit the parametric down-conversion rate~\cite{Bengtsson2018a}. In microwave experiments such as ours, where nonlinearities induced by Josephson junctions are relatively much larger than in optics, the second process dominates the first one and is thus responsible for parametric oscillation stabilization~\cite{Wustmann2017}.

The Kerr nonlinearities are generally described by an extra term in the Hamiltonian, which can be written in the rotating wave approximation \cite{Flurinthesis}
\begin{equation}
\hat{H}_{K} = \hbar \left[\chi_{a}(\hat{a}^\dagger)^2 \hat{a}^2+\chi_{b}(\hat{b}^\dagger)^2 \hat{b}^2+\chi_{ab}\hat{a}^\dagger \hat{a}\hat{b}^\dagger \hat{b}\right]
\label{H4}
\end{equation}
where $\chi_a$ and $\chi_b$ are self-Kerr coefficients of the two resonators and $\chi_{ab}=4\sqrt{\chi_a\chi_b}$ is their cross-Kerr coefficient. They effectively modify the oscillator frequencies by a detuning $\chi_a(\hat{a}^\dagger \hat{a})+\chi_{ab}(\hat{b}^\dagger \hat{b})$ for resonator $a$ and $\chi_b(\hat{b}^\dagger \hat{b})+\chi_{ab}(\hat{a}^\dagger \hat{a})$ for resonator $b$.

As parametric oscillation is stabilized, the resulting outgoing field amplitudes $a_{out}$ and $b_{out}$ exhibit similar properties as in previously studied spatially degenerate Josephson amplifiers~\cite{Wilson2010, Bengtsson2018a}. Performing repeated heterodyne quadrature measurements of the outgoing field amplitudes reveals a statistics of quadratures that is characteristic of parametric oscillation (see Figs. 6 and 7). As previously demonstrated~\cite{Bengtsson2018a}, we observe that non-degenerate parametric oscillations occur with an arbitrary phase, as expected from the highly degenerate state of the system.

\section{Injection locking}

\begin{figure}
\includegraphics[scale=0.45]{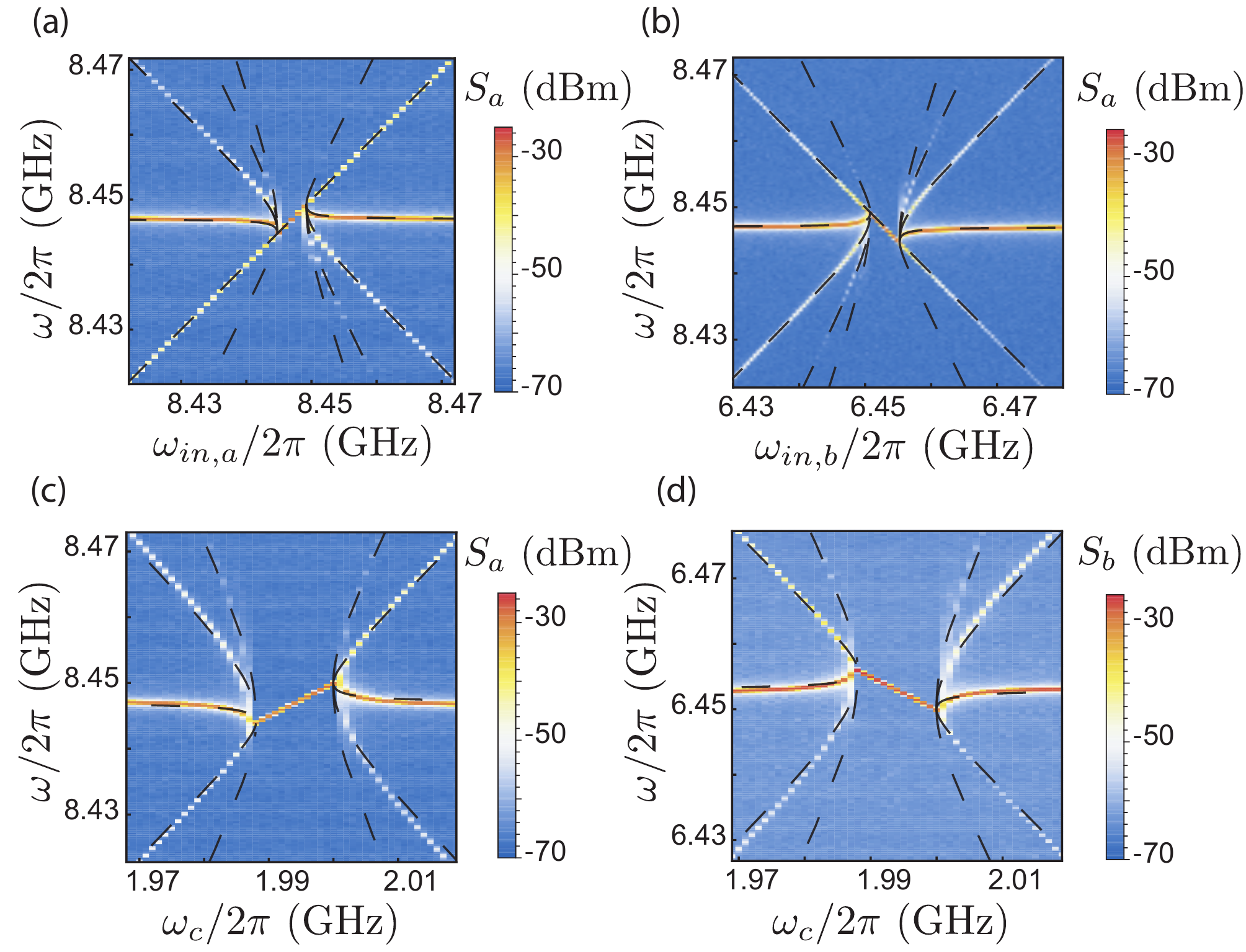}
\caption{(a)-(b) Measured power spectral density $S_a(\omega)$ emitted on mode $a_{out}$, as a function of emission frequency $\omega$ and frequency $\omega_{in,a}$ (resp. $\omega_{in,b}$) of the tone injected on the mode $a$ (resp. $b$). The signal is injected at a fixed power $P_{in} = -26$ dBm both for the mode $a$ and $b$. Black dashed lines correspond to Adler's theory (see text). (c)-(d) Measured power spectral densities $S_a(\omega)$ and $S_b(\omega)$ emitted from the modes $a_{out}$ and $b_{out}$, as a function of emission frequency $\omega$ and frequency $\omega_{c}$ of the tone injected on the common mode at fixed power $P_c = -26$ dBm. For all measurements the pump frequency is fixed at $\omega_p = 2 \pi \times 14.9$ GHz and power $P_p = -16$ dBm.}
\label{spectra}
\end{figure}

The standard approach to suppress phase indeterminacy of parametric oscillations is to use injection locking. It consists in injecting a small signal at the resonance frequency of the oscillator hence breaking the phase degeneracy of the system. The oscillations lock on the phase of the injected signal, which consequently narrows the linewidth of the emissions. In Fig.~\ref{spectra}(a), we show that this usual scheme can be applied to our device by driving the mode $a$ at a frequency $\omega_{in,a}$, close to its natural resonance frequency $\omega_a$ (Fig.~\ref{spectra}(a)). Here, the pump power is fixed to $P_p=-16~\mathrm{dBm}$ (referred to the input of the dilution refrigerator) so that the cooperativity is set to $C\approx1$. With a constant input drive power at $\omega_{in,a}$, we measure the spectral noise power $S_a(\omega)$ in the output mode $a_{out}$ as a function of $\omega_{in,a}$. At low enough detuning $\omega_{in,a}-\omega_{a}$, a single peak can be seen in the spectrum and is localized at $\omega_{in,a}$ (Fig.~\ref{spectra}(a)), hence indicating that the oscillator frequency is pulled by the input tone and gets locked to it. This process works for input frequencies within some injection locking range $|\omega_{in,a}-\omega_a|<\Delta\omega_{in}/2$. Beyond that range, the emission spectrum consists of a series of peaks that match Adler's theory \cite{Adler1973} for injection locking since they are localized at
\begin{equation}
\omega_n[\omega_{in,a}] = \omega_{in,a} + (n+1) \omega_{beat}[\omega_{in,a}]
\label{omegan}
\end{equation}
for an arbitrary integer index $n$ where the beating frequency $\omega_{beat}$ is defined as \cite{Armand1969}
\begin{equation}
\omega_{beat}[\omega] =(\omega_a - \omega)\sqrt{1- \left(\frac{\Delta \omega_{in}}{2(\omega_a - \omega)}\right)^2}.
\label{omegab}
\end{equation}
The resulting frequencies $\omega_n[\omega_{in,a}]$, shown as dashed lines in Fig.~\ref{spectra}, match precisely with the position of the peaks in the emission spectra of the circuits. The peak $n = 0$ corresponds to the pulled oscillator frequency, which becomes equal to its natural frequency when $|\omega_{in,a}-\omega_a|\gg\Delta\omega_{in}$ and deviates significantly as $\omega_{in, a}$ approaches $\omega_a$. In agreement with the theory, we observe experimentally that $\omega_0$ evolves continuously as we sweep the injection frequency and matches $\omega_{in,a}$ in the injection locking range. The peak corresponding to $n = -1$ is exactly at the injection frequency for all $\omega_{in,a}$. Other values of $n$ can be observed and correspond to higher order frequency distortion sidebands. 

Since, for non-degenerate parametric oscillations, the phases of the oscillations are correlated in resonators $a$ and $b$, it is sufficient to inject a signal in only one of them to suppress simultaneously their phase indeterminacy. As a consequence, by injecting a signal at frequency $\omega_{in,b}\approx \omega_b$ in $b$, we observe the same injection locking phenomena than before when looking at the emission spectrum of resonator $a$ (Fig.~\ref{spectra}(b)). One striking difference is that, within the injection locking range, the frequency of the emission peak is given by $\widetilde{\omega}_{in,a} = \omega_{p} - \omega_{in,b}$ rather than $\omega_{in,a}$, and therefore decreases as we increase the frequency of the injected signal. One advantage of this two-mode injection locking is that the injected signal can be spatially and spectrally separated from the natural frequency of the oscillator. This could be useful for certain applications, where for example one of the resonators has to remain isolated such as in a quantum memory \cite{Flurin2015}. Beyond the locking range, the emission spectrum frequency peaks are localized at $\omega_n[\widetilde{\omega}_{in,a}]$ according to Eq.~(\ref{omegan}). The measured spectral peaks match well these frequencies (dashed lines in Fig.~\ref{spectra}(b)), so that their frequencies are also captured by Adler's theory when replacing $\omega_{in,a}$ by $\omega_{p} - \omega_{in,b}$.

\begin{figure}
\includegraphics[scale=0.4]{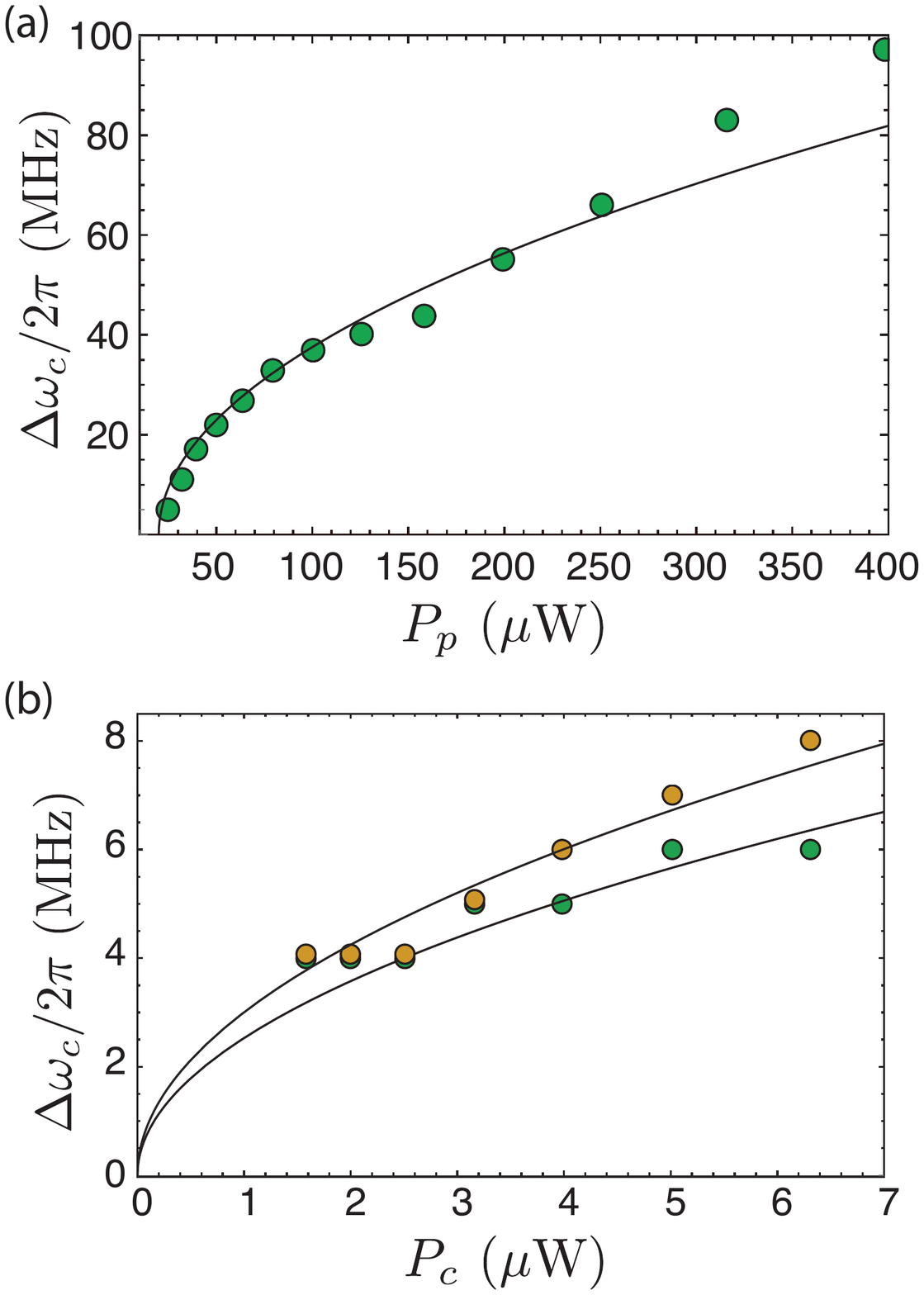}
\caption{(a) Parametric locking range $\Delta\omega_{c}$ measured on the oscillator $a$ as a function of pump power $P_p$. The pump frequency is set to $\omega_p = 2 \pi \times 14.95$ GHz and the parametric locking tone is injected at $\omega_{c}$ for a constant power $P_{c} = -20$ dBm. Full black line corresponds to $\Delta \omega_{c} \propto \sqrt{P_{p}})$. (b) Parametric locking range measured on the modes $a$ (green dots) and $b$ (orange dots) as a function of locking signal power $P_{c}$ at $\omega_{c}$. The pump frequency is set to $\omega_p = 2 \pi \times 14.95$ GHz and power $P_p = -16$ dBm. Full black lines correspond to $\Delta \omega_{c} \propto \sqrt{P_{c}})$.}
\label{lockrange}
\end{figure}

\begin{figure*}
\includegraphics[scale=0.85]{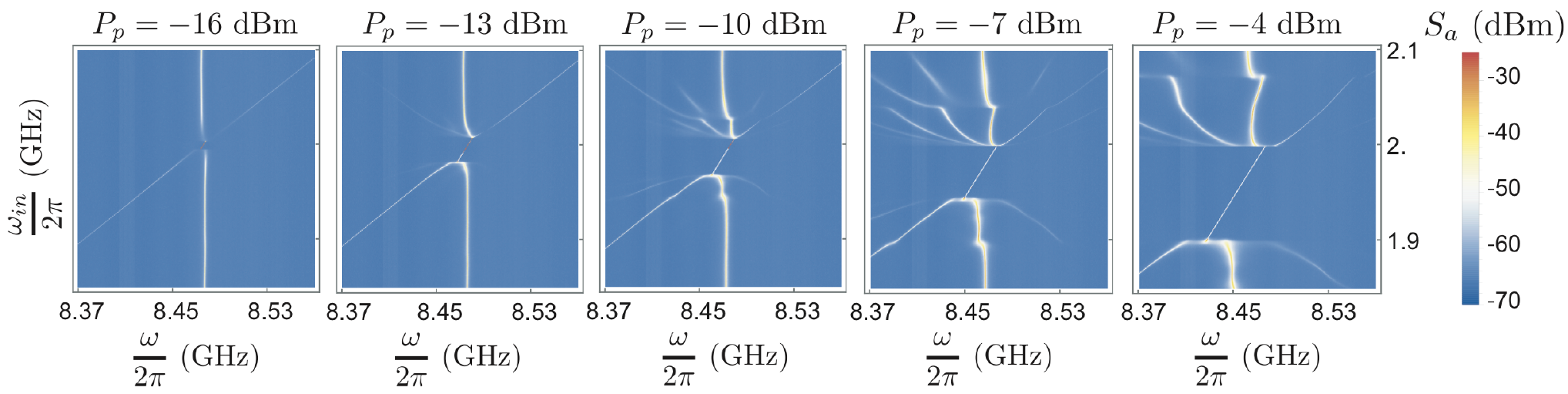}
\caption{Measured power spectral density $S_a(\omega)$ of mode $a_{out}$ as a function of $\omega$ and $\omega_{c}$ for different pump powers $P_p$. The pump frequency is set to $\omega_p = 2 \pi \times 14.952$ GHz and the locking signal is injected at a constant power $P_c = -20$ dBm.}
\label{film}
\end{figure*}

\section{Parametric locking}

We demonstrate a new strategy to suppress phase indeterminacy, which relies on what we call "parametric locking" instead of injection locking. Since the sum of the phases of the two modes $\theta_a+\theta_b$ is fixed~\cite{Wustmann2017} when the system is pumped at $\omega_a+\omega_b$, one only needs a single extra constraint to determine the phases of the outgoing field unequivocally. This can be obtained by pumping the circuit with a parametric drive close to $\omega_a-\omega_b$, which connects the fields in $a$ and $b$ through processes of conversion such that $\theta_a-\theta_b$ becomes a constant~\cite{Abdo2013}. In practice, we drive the common mode of the Josephson mixer~\cite{Flurinthesis} at $\omega_p = \omega_a + \omega_b$ with an amplitude $p$ above the parametric oscillation threshold, and simultaneously at $\omega_{c} \approx \omega_a - \omega_b$ with an amplitude $c$. This simultaneous pumping can be used in other power regimes to realize a circulator~\cite{Sliwa2015, Chien2019} or simulate ultrastrong coupling regime between oscillators~\cite{Markovic2018}. 

Due to this additional drive at $\omega_c$, the three-wave mixing Hamiltonian acquires extra terms in the RWA
\begin{equation}
\hbar \chi ( c \hat{a}^\dagger \hat{b} + c^* \hat{a}\hat{b}^\dagger),
\end{equation}
which account for the conversion of photons between resonators $a$ and $b$. They result in a parametric locking similar to injection locking within a range $\Delta\omega_{c}$, in which the frequency of emission is given by $(\omega_p + \omega_{c})/2$ instead of $\omega_{in,a}$ for the mode $a$ and $(\omega_p - \omega_{c})/2$ instead of $\omega_{in,b}$ for the mode $b$. Moreover, beyond this range, the spectrum is also composed of a series of peaks, whose frequencies are properly described by Adler's theory provided that in Eq.~\eqref{omegan} we use an effective injection locking frequency $\widetilde{\omega}_{in,a} = (\omega_p + \omega_{c})/2$ for the mode $a$ and $\widetilde{\omega}_{in,b} = (\omega_p - \omega_{c})/2$ for the mode $b$.

This behavior is shown in Fig.~\ref{spectra}(c) and (d), where we have measured power spectral densities $S_a(\omega)$ and $S_b(\omega)$ of the radiation emitted respectively from modes $a_{out}$ and  $b_{out}$, as a function of the frequency $\omega_{c}$ of the parametric drive. The spectra obtained by parametric locking are similar to those obtained by injection locking scheme (Figs.~\ref{spectra}(a,b)). However a couple of striking differences can be observed. Since the effective locking frequency is now given by $(\omega_p \pm \omega_c)/2$, the frequency of the emission peak within the locking range shows a slope $\pm \frac{1}{2}$ as a function of $\omega_{c}$ for modes $a$ and $b$ respectively. This technique of parametric locking allows to suppress phase indeterminacy by using only tones that are far from the resonant frequencies of the oscillators. This procedure could be useful for certain applications when one wants the locking signal to remain far from the spectral range of interest. It could also be used in situation where the frequencies of two oscillators are close but outside of the technically accessible range. Outside of the locking range, the emission peak frequencies are properly described using Adler's theory with the effective locking frequencies (dashed lines in Figs.~\ref{spectra}(c,d)).

Parametric locking can be made to operate on a larger frequency range by increasing the pump power $P_p$ at $\omega_p$ away from the parametric oscillation threshold. It can be shown using Langevin equations (see Appendix A) that at low pump powers, we expect $\Delta\omega_{c}$ to be proportional to the pump amplitude $|p|$ and thus to $\sqrt{P_{p}}$. It is another qualitative difference compared to the case of injection locking where $\Delta\omega_{in}$ is proportional to $\sqrt{|p|}$ and thus to $P_{p}^{\frac{1}{4}}$. As shown in Fig. 3(a), we observe that the size of the locking range indeed follows the expected behavior as a function of pump power. At larger pump power $p$, higher order terms start to contribute significantly and the locking range has a more complex evolution. 

We also observe a power dependence of the parametric locking range with the power of the injected parametric signal. It is well known for standard injection locking \cite{Liu2015} that the locking range is proportional to the amplitude of the injected signal and is therefore a square root function of the injected power $P_{in}$ as predicted by Adler's theory. We observe that for the parametric locking the locking range increases similarly with the strength of the injected parametric signal (Figure~\ref{lockrange}(b)) suggesting that a stronger injection signal is favorable to suppress phase indeterminacy in resonators $a$ and $b$. Note that for these measurements as well as those in Fig.~\ref{spectra}, the injection signal powers were chosen so that the frequency of the peak in the middle of the locking range is the same for all experiments in order to ease comparisons.

Finally, we explore the parametric locking behavior of our circuit well beyond the parametric threshold by increasing the pump power $P_p$. This reveals qualitative deviations to Adler's theory as can be seen in the measured spectra of Fig.~\ref{film}. At powers $P_p>-10~\mathrm{dBm}$ (referred to the input of the dilution refrigerator), the bifurcation like features~\cite{Richy95} appear in the emission peak frequency dependence due to the importance of microscopic multiphotonic processes involving more than three photons. These features are reminiscent of chaotic behavior~\cite{Lugiato1988} and should be studied further.

\section{Conclusion}

We have demonstrated injection locking and parametric locking techniques for coupled non-linear oscillators that make use of a three-wave mixing interaction. The geometry of the Josephson circuit we use, based on two spatially and spectrally separated resonators, provides original approaches to suppress the inherent phase indeterminacy of parametric oscillators and lock the field emitted from the resonators. These new techniques could be useful for applications in which standard injection locking is not possible. This work illustrates how the strong coupling of superconducting circuits to microwave modes enables to design and use a variety of new nonlinear effects that will be instrumental for quantum information applications such as quantum error correction, non-reciprocal devices and detectors. 
\begin{acknowledgments}
We thank Zaki Leghtas, Ananda Roy, Thierry Dauxois, Quantronics group and Michel Devoret for fruitful interactions over the course of this project. Nanofabrication has been made within the consortium Salle Blanche Paris Centre. This work was supported by the EMERGENCES grant QUMOTEL of Ville de Paris, by the French Agence Nationale
de la Recherche (GEARED project No. ANR-14-CE26-0018).
\end{acknowledgments}

\appendix

\section{Parametric oscillation}

In our experiment, we measure emission spectra of a Josephson circuit in the parametric oscillation regime. The circuit is driven by a pump at an amplitude larger than those used in linear parametric amplification regime. The threshold pump amplitude is defined by the cooperativity $C$ equal to $1$ and is thus given by $|\chi p^{th}|^2 = \frac{ \kappa_a \kappa_b}{4}$.

We detect the threshold by measuring the distribution of the field quadratures for different pump amplitudes. The signal coming out from the mode $a$ (resp. $b$) is amplified and its two quadratures $X_a$ and $P_a$ (resp. $X_b,P_b$) are determined by heterodyne measurement. In practice, the signal emitted from each of the modes $a$ and $b$ is amplified and mixed with an external local oscillator with a small (few tens of MHz) detuning $\delta \omega$ from the mode frequencies. The output voltage of the mixer is then digitized using a 250 MHz bandwidth acquisition board. We demodulate the signals at $\delta\omega$ over a finite time window in order to determine the two quadratures $X$ and $P$. It is possible to calibrate these quadratures in meaningful units. If a coherent state $|\alpha\rangle$ is stabilized in the mode $a$, the calibration is such that, on average, $\langle X_a\rangle=\mathrm{Re}(\alpha)$ and $\langle P_a\rangle=\mathrm{Im}(\alpha)$. With this choice, the quadratures $X$ and $P$ are dimensionless.

\begin{figure}[!h]
\includegraphics[scale=0.33]{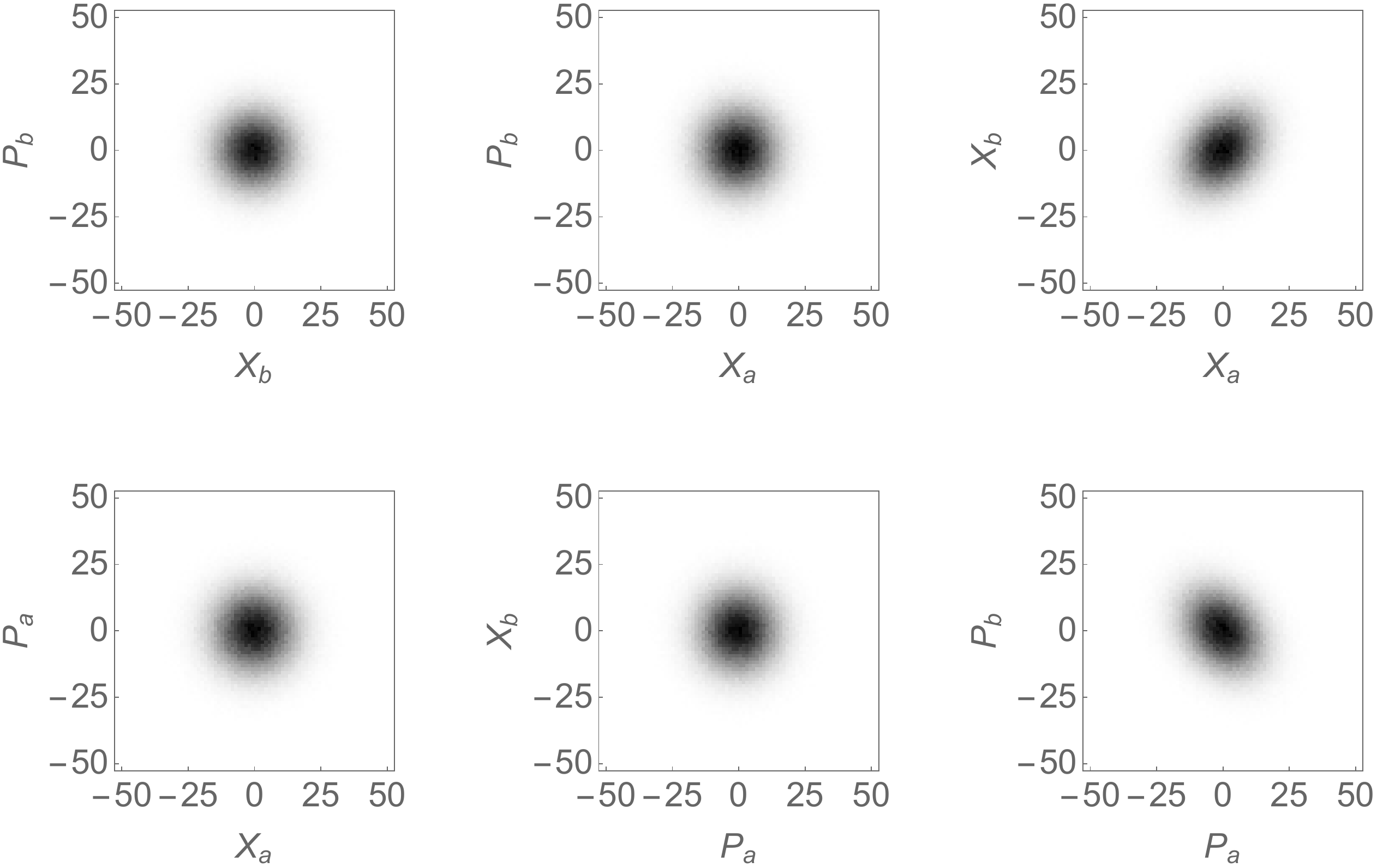}
\caption{Field quadrature histograms of the output fields measured in the
linear amplification regime. They correspond to the emission of a two-mode squeezed state by amplification of vacuum fluctuations, as can be inferred from the histograms shown along quadratures of different modes (see \cite{Flurin2015} for details).}
\label{SP1}
\end{figure}

In the parametric amplification regime, when the modes $a$ and $b$ are not driven, the output field is described by a Gaussian distribution centered around the origin. The variance of a single mode $a$ or $b$ is that of the vacuum fluctuations amplified by the Josephson mixer and with the thermal noise added by the chain of amplifiers, as shown in Fig.~\ref{SP1}. The apparent two-mode squeezing in the histograms on $X_a$--$X_b$ and $P_a$--$P_b$ originates from to the vacuum two-mode squeezed state emitted by the device~\cite{Flurin2012}.

\begin{figure}[!h]
\includegraphics[scale=0.33]{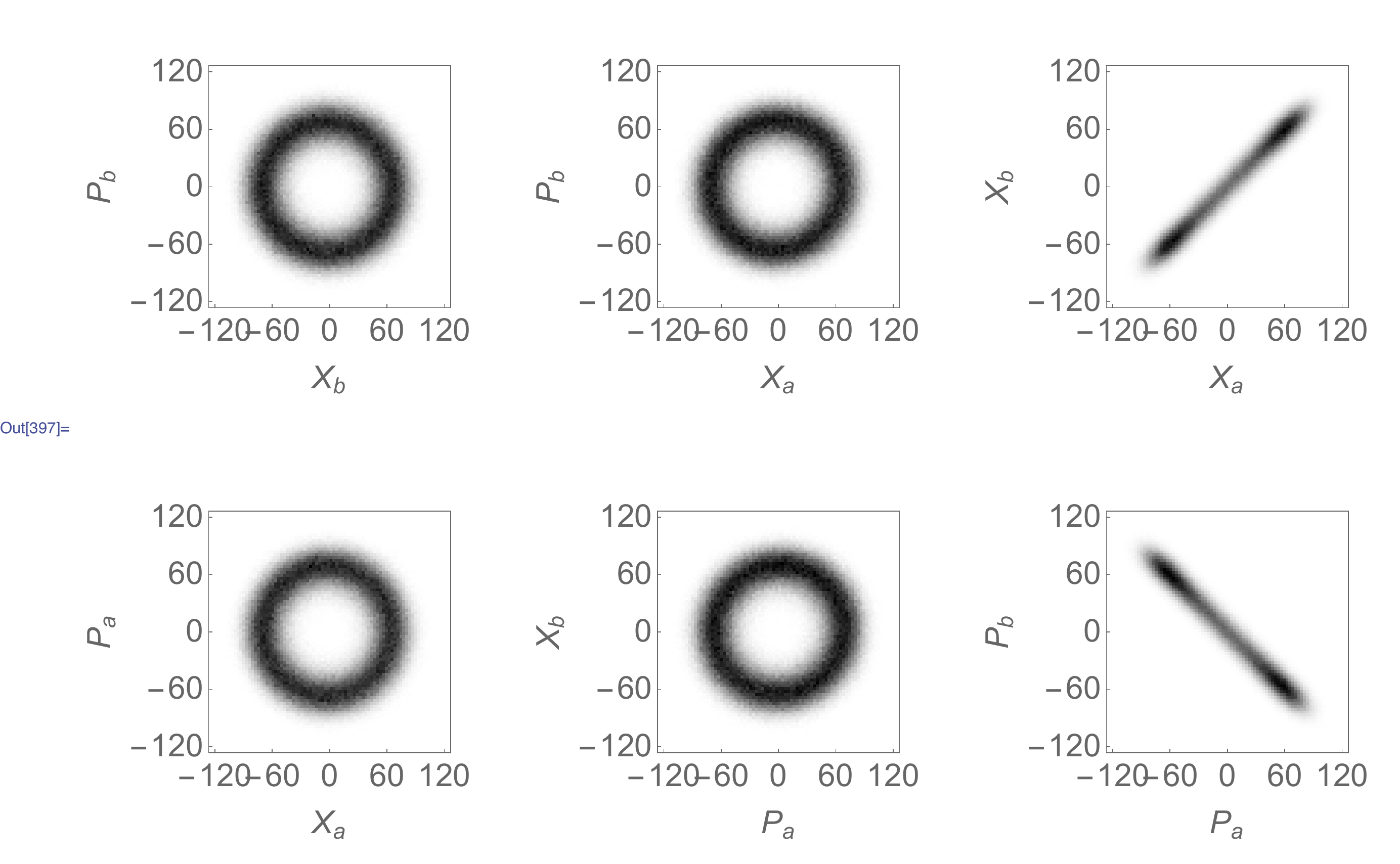}
\caption{Distribution of measurement outcomes in the
parametric self-oscillation regime at cooperativity $C=1.3$. The phase of the pump was tuned such that strong correlations are visible in the $X_a$--$X_b$ and $P_a$--$P_b$ planes.}
\label{SP2}
\end{figure}

Beyond threshold the self-sustained parametric oscillation state is characterized by complex amplitudes given by
\begin{align*}
\begin{cases}
a = a_0 e^{i \theta_a}\\
b = b_0 e^{i \theta_b}
\end{cases}
\end{align*}
and highly degenerate phases $\theta_a$ and $\theta_b$ of the fields~\cite{Bengtsson2018a}. This regime is shown in Fig.~\ref{SP2}. Note that, while a minimum of potential energy is reached for any phase $\theta_a$ or $\theta_b$, the sum between the phases is constrained and given by the phase of the pump up to an offset that depends on cooperativity. In Fig.~\ref{SP2}, we chose the phase of the pump to highlight this phase constraint in the $X_a$--$X_b$ and $P_a$--$P_b$ phase spaces.

\begin{figure}[!h]
\includegraphics[scale=0.28]{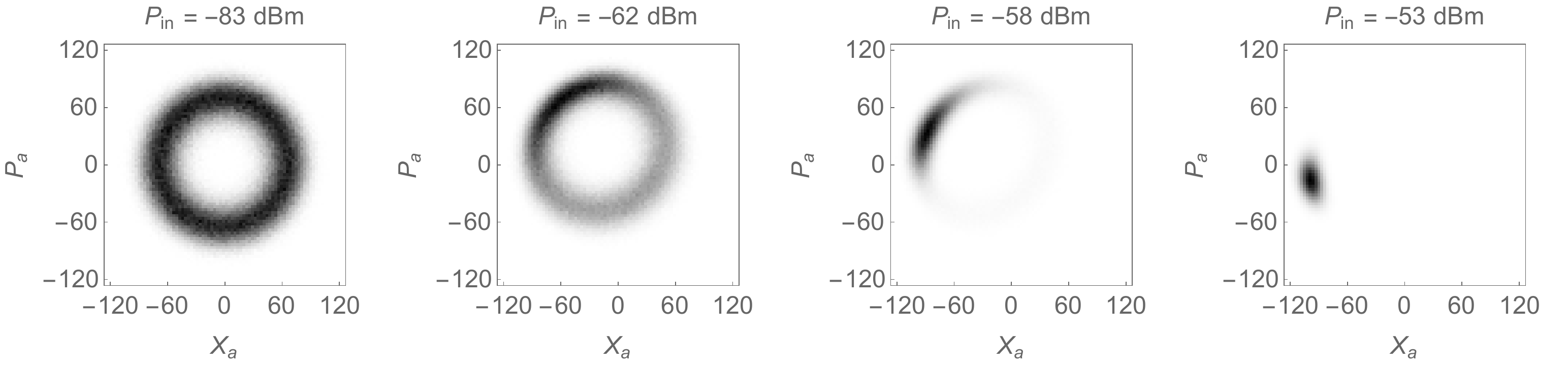}
\caption{Measured quadrature statistics for the mode $a$ for the pump power fixed in the self-oscillation regime and for four different powers of the small signal injected on resonance into the mode $a$. Field quadrature axes are calibrated in number of photons.}
\label{SP3}
\end{figure}

We then inject a weak tone at frequency $\omega_{in}$ on resonance in the mode $a$ and measure the
quadrature statistics for different powers of injected signal. Histograms shown in Fig.~\ref{SP3} demonstrate the transition from the phase unlocked to the phase locked regime. A global rotation from one histogram to the next corresponds to a slowly drifting phase in the detection setup.

\section{Injection locking and parametric locking range calculated using Langevin equation formalism}

\subsection{Injection locking}

When the circuit is pumped at $\omega_p = \omega_a + \omega_b$, it can be described by the classical Langevin equations
\begin{align}
    \begin{cases}
    \begin{array}{rl}
    \frac{da}{dt} =  -i \omega_a a - i \chi p b^\ast - \Gamma_a a + i \chi_a |a|^2 a & + i \chi_{ab} |b|^2 a \\
    & + \sqrt{2 \Gamma_a}a_{in} \\
    \end{array}\\
    \begin{array}{rl}
      \frac{db}{dt} = -i \omega_b b - i \chi p a^\ast - \Gamma_b b + i \chi_b |b|^2 b & + i \chi_{ab} |a|^2 b\\
      &+ \sqrt{2 \Gamma_b}b_{in}\\
    \end{array}
    \end{cases}
\label{Langevin_IL}
\end{align}
where $\Gamma_{a,b} = \kappa_{a,b}/2$ and $\kappa_{a,b}$ are the dissipation rates of the modes of amplitude $a$ and $b$, and $p=|p|e^{-i\omega_p t}$ is the amplitude of the pump applied on the common mode. We have included fourth order Kerr terms necessary for the stabilization of parametric oscillation.

On top of these equations, the fields, which can have complex amplitudes, obey the following boundary conditions 
\begin{align}
    \begin{cases}
\sqrt{2\Gamma_a}a=a_{in}+a_{out}\\
\sqrt{2\Gamma_b}b=b_{in}+b_{out}
    \end{cases}
\end{align}
In the parametric oscillation regime and beyond threshold, the number of photons grows rapidly in the resonators ($n_a\gg1$ and $n_b\gg1$), such that the outgoing power is much larger than the incoming one. We then have $\sqrt{2\Gamma_a}a\approx a_{out}$ and $\sqrt{2\Gamma_b}b\approx b_{out}$. Because of energy conservation, the number of outgoing photons must be the same for resonators $a$ and $b$ and therefore $|a_{out}|\approx|b_{out}|$. It is therefore straightforward to show that
\begin{equation}
\Gamma_a n_a\approx\Gamma_b n_b,
\label{photon_eq}
\end{equation}
if $n_a=|a|^2$ and $n_b=|b|^2$ are the numbers of photons in each mode.

For injection locking to occur, one injects a small signal $a_{in} = |a_{in}| e^{-i \omega t}$ in the resonator $a$, whereas $b_{in} = 0$, such that the fields $a$ and $b$ acquire well defined phases and frequencies. We can therefore write $a=\tilde{a}e^{-i\omega t}$ and $b=\tilde{b}e^{-i(\omega_p-\omega) t}$, where the complex amplitudes $\tilde a$ and $\tilde b$ do not depend on time. Using these expressions in Eq.~\eqref{Langevin_IL}, we find the following expression for the field in the resonator $a$
\begin{widetext} 
\begin{equation}
a=\frac{\sqrt{2\Gamma_{a}}a_{in}}{-i\left(\Delta\omega+\chi_{a}n_a+\chi_{ab}n_b\right)+\Gamma_{a}-\frac{\left|\chi p\right|^{2}}{i\left(-\Delta\omega+\chi_{b}n_b+\chi_{ab}n_a\right)+\Gamma_{b}}} ,
\label{field}
\end{equation}
\end{widetext} 
where we have defined $\Delta\omega=\omega-\omega_a$.

For a large field to arise in the resonator ($|a|\gg 1$) when a vanishing injection tone amplitude $|a_{in}|\rightarrow 0$ is applied, the  denominator of the previous expression must become smaller than $\sqrt{\Gamma_a}|a_{in}|$. Therefore, when injection locking happens, one can write
\begin{widetext}
\begin{equation}
\left[-i\left(\Delta\omega+\chi_{a}n_a+\chi_{ab}n_b\right)+\Gamma_{a}\right]\left[i\left(-\Delta\omega+\chi_{b}n_b+\chi_{ab}n_a\right)+\Gamma_{b}\right]-\left|\chi p\right|^{2}\rightarrow 0
\end{equation}
\end{widetext}
The real part of this expression gives the condition for injection locking to occur
\begin{equation}
|\chi p|^2\approx\Gamma_a\Gamma_b+(\Delta\omega+\chi_{a}n_a+\chi_{ab}n_b)(-\Delta\omega+\chi_{b}n_b+\chi_{ab}n_a),
\label{condition1}
\end{equation}
while the imaginary part gives the relation between the photon number and the frequency shift with respect to the natural frequency of the oscillator
\begin{equation}
\Gamma_b(\Delta\omega+\chi_{a}n_a+\chi_{ab}n_b)\approx\Gamma_a(-\Delta\omega+\chi_{b}n_b+\chi_{ab}n_a).
\label{condition2}
\end{equation}
Using the condition \eqref{photon_eq} that $\Gamma_a n_a\approx\Gamma_b n_b$, one respectively obtains from Eq.~\eqref{condition1} and ~\eqref{condition2} 
\begin{equation}
|\chi p|^2=\Gamma_a\Gamma_b+(\Delta\omega+\tilde\chi_{a}n_a)(-\Delta\omega+\tilde\chi_{b}n_a)
\label{condition1bis}
\end{equation}
and
\begin{equation}
\Gamma_b(\Delta\omega+\tilde\chi_{a}n_a)=\Gamma_a(-\Delta\omega+\tilde\chi_{b}n_a)
\label{condition2bis}
\end{equation}
where we have defined $\tilde{\chi}_a = \chi_a + \chi_{ab} \Gamma_a/ \Gamma_b$, $\tilde{\chi}_b = \chi_{ab} + \chi_{b} \Gamma_a/ \Gamma_b$. Equations \eqref{condition1bis} and \eqref{condition2bis} have a single solution $(\Delta\omega_0,n_{a0})$, which means that the frequency range over which the locking can occur is infinitely small ($\Delta\omega_{in}=\Delta\omega_{max}-\Delta\omega_{min}\approx0$).

For larger injection tone amplitude, \emph{i.e}. $|a_{in}|\sim\sqrt{\Gamma_a}$, but still under the condition $\Gamma_a n_a\gg|a_{in}|^2$, the square modulus of Eq.~\eqref{field} combined with Eq.~\eqref{photon_eq} gives
\begin{equation}
    n_a=\left|\frac{\sqrt{2\Gamma_{a}}a_{in}}{-i\left(\Delta\omega+\tilde\chi_{a}n_a\right)+\Gamma_{a}+\frac{\left|\chi p\right|^{2}}{i\left(-\Delta\omega+\tilde\chi_{b}n_a\right)+\Gamma_{b}}}\right|^2
    \label{ellipselike}
\end{equation}
which has an ensemble of solutions forming an ellipse-like shape centered at $(\Delta\omega_0,n_{a0})$ in the $\left\{\Delta\omega,n_a\right\}$ phase space, and whose major axis is given by Eq.~\eqref{condition1bis}. We can therefore find an approximate expression for the maximum and minimum frequency shift $\Delta\omega$ along this axis by combining Eq.~\eqref{ellipselike} and \eqref{condition1bis}. This leads to
\begin{equation}
\begin{array}{rcl}
    n_a&\times&\left[\Gamma_b(\Delta\omega+\tilde\chi_{a}n_a)-\Gamma_a(-\Delta\omega+\tilde\chi_{b}n_a)\right]^2]=  \\
     && 2\Gamma_a |a_{in}|^2 \left[\Gamma_b^2+(\Delta\omega-\tilde\chi_b n_a)^2\right]
\end{array}
\end{equation}
and then by introducing $\delta\omega=\Delta\omega-\Delta\omega_0$ and $\delta n_a=n_a-n_{a0}$, we obtain the relation between the resonator population $\delta n_a$ for an injection amplitude $a_{in}$ and
the frequency shift $\delta\omega$
\begin{equation}
\begin{array}{rcl}
    n_a&\times&\left[\Gamma_b(\delta\omega+\tilde\chi_a \delta n_a)-\Gamma_a(-\delta\omega+\tilde\chi_b \delta n_a)\right]^2=  \\
     && 2\Gamma_a |a_{in}|^2\left[\Gamma_b^2+(\Delta\omega-\tilde\chi_b n_a)^2\right],
\end{array}\label{relation}
\end{equation}
where we have used the fact that $(\Delta\omega_0,n_{a0})$ is solution of Eq.~\eqref{condition2bis}.

Eq.~\eqref{relation} can be simplified by introducing the three constants
\begin{align}
    \begin{cases}
    \alpha = (\Gamma_a + \Gamma_b) + (\Gamma_b \tilde{\chi}_a - \Gamma_a \tilde{\chi}_b) \times \frac{2 \Delta\omega_0 - (\tilde{\chi}_b - \tilde{\chi}_a) n_{a0}}{(\tilde{\chi}_b - \tilde{\chi}_a) \Delta\omega_0+ 2 \tilde{\chi}_a\tilde{\chi}_b n_{a0}} \\
    \beta = \frac{\tilde{\chi}_b \Gamma_a - \tilde{\chi}_a \Gamma_b }{ \Gamma_a + \Gamma_b} - \tilde{\chi}_b \\
    \gamma = |\frac{\Gamma_a + \Gamma_b}{\tilde{\chi}_a + \tilde{\chi}_b }| \times \sqrt{\frac{1}{\Gamma_a \Gamma_b}}
    \end{cases}
\end{align}
which linearly relates $\delta\omega$ to $\delta n_a$, $\Delta\omega_0$ to $n_{a0}$ and $n_{a0}$ to $|\chi p|$ in the following way
\begin{align}
	\begin{cases}
\Gamma_b(\delta\omega+\tilde{\chi}_{a}\delta n_{a})-\Gamma_a(-\delta\omega+\tilde\chi_b \delta n_a)=\alpha\times\delta\omega\\
\Delta\omega_{0}-\tilde{\chi}_{b}n_{a0}=\beta\times n_{a0}\\
n_{a0}=\gamma\times\left|\chi p\right|.
	\end{cases}
\end{align}
given that the number of photons in the resonator remains very large ($\Gamma_a\Gamma_b\ll|\chi p|^2$ and $\Gamma_b\ll\tilde\chi_b n_a$). The constants $\beta$ and $\gamma$ are obtained using the fact that $(\Delta\omega_0,n_{a0})$ is solution of Eq.~\eqref{condition1bis} and Eq.~\eqref{condition2bis}, while $\alpha$ is obtained by linearization of Eq.~\eqref{condition1bis} to the first order in $\delta\omega$ and $\delta n_a$.

At the lowest order in $\delta\omega$, $\delta n_a$ and $|a_{in}|$ and for $\Gamma_b\ll\tilde\chi_a n_{a0}$, Eq.~\eqref{relation} then becomes
\begin{equation}
\gamma\times\left|\chi p\right|\alpha^2\delta\omega^2=2\Gamma_a|a_{in}|^2\beta^2 (\gamma\times\left|\chi p\right|)^2
\end{equation}
which gives two symmetric solutions $\pm\delta\omega$ such that the locking range  $\Delta\omega_{in}=2\delta\omega$ is given by
\begin{equation}
    \Delta \omega_{in} \approx 2 \sqrt{2\Gamma_a} \sqrt{\frac{\beta^2 \gamma}{\alpha^2}} |a_{in}| \sqrt{|\chi p|},
\label{locking range}
\end{equation}

Expression \eqref{locking range} shows that the locking range grows as the square root of the injected power $\sqrt{P_{in}}=|a_{in}|$ and follows a power law with the pump power $\sqrt{|p|}=P_p^{1/4}$ as mentioned in Fig.~\ref{lockrange}.

\subsection{Parametric locking}

In the case of the parametric locking, the circuit is simultaneously pumped at $\omega_p = \omega_a+\omega_b$ and $\omega_c = \omega_a - \omega_b$, and $a_{in} = b_{in} = 0$. The Langevin equations thus read
\begin{align}
    \begin{cases}
    \frac{da}{dt} = -i \omega_a a - i \chi p b^\ast - \Gamma_a a + i \chi_a |a|^2 a + i \chi_{ab} |b|^2 a - i \chi c b \\
      \frac{db}{dt} = -i \omega_b b - i \chi p a^\ast - \Gamma_b b + i \chi_b |b|^2 b + i \chi_{ab} |a|^2 b - i \chi c^\ast a
    \end{cases}
\end{align}
where $p=|p|e^{-i(\omega_p t+\varphi_p)}$ and $c=|c|e^{-i(\omega_c t+\varphi_c)}$ are amplitudes of the stiff pumps applied to the common mode of the circuit.

The solutions for which parametric locking occurs can be written as before as $a=\tilde a e^{-i\omega_1 t}$ and $b=\tilde b e^{-i\omega_2 t}$. Stationary solutions, where $\tilde a$ and $\tilde b$ do not depend on time, only exist if
\begin{equation}
\omega_1=\frac{\omega_p+\omega_c}{2}\text{ and } \omega_2=\frac{\omega_p-\omega_c}{2}.
\end{equation}
Injecting these solutions into the Langevin equations leads straightforwardly to the expression
\begin{widetext}
\begin{equation}
\frac{a}{a^{*}}=\frac{2i\chi^2 pc\left(\Delta\omega-\tilde{\chi}_{b}n_{a}\right)/\left[\left(\Delta\omega-\tilde{\chi}_{b}n_{a}\right)^{2}+\Gamma_{b}^{2}\right]}{-i\Delta\omega+\Gamma_{a}-i\tilde{\chi}_{a}n_{a}-\frac{\left|\chi p\right|^{2}}{\Gamma_{b}-i\left(\Delta\omega-\tilde{\chi}_{b}n_{a}\right)}{+\frac{\left|\chi c\right|^{2}}{\Gamma_{b}+i\left(\Delta\omega-\tilde{\chi}_{b}n_{a}\right)}}}.
\label{paramlock}
\end{equation}
\end{widetext}
where we have taken $\omega_1=\omega$, $\omega_2=\omega_p-\omega$ and define $\Delta\omega=\omega-\omega_a$.

Similarly to injection locking in the case of vanishing input amplitude $a_{in}$, this expression has a single solution $(\Delta\omega_0,n_{a0})$ if $c\rightarrow 0$.

For a finite pump amplitude $c$, the solutions also form an ellipse in the $\left\{\Delta\omega,n_a\right\}$ phase space whose center is given by $(\Delta\omega_0,n_{a0})$ and major axis by the solution of Eq.~\eqref{condition1bis}. Similarly to injection locking, we can find an approximate expression for the maximum and minimum frequency shifts $\Delta\omega$ along this axis combining Eq.~\eqref{condition1bis} and \eqref{paramlock} and using the fact that in the experiment $|p|\gg|c|$ and that numbers of photons $n_a$ and $n_b$ are large. The expression \eqref{paramlock} thus simplifies to
\begin{equation}
\frac{a}{a^*}\approx\frac{i2\chi^2 pc}{\alpha \delta \omega},
\end{equation}
where $\alpha$ is the same constant than for injection locking and $\delta\omega=\Delta\omega-\Delta\omega_0$.

We thus find the parametric locking range to be
\begin{equation}
    \Delta \omega_{c} = 4\chi^2 \frac{|pc|}{\alpha},
\end{equation}
which shows that the locking range grows as $\sqrt{P_p}$ and $\sqrt{P_c}$. The two pumps therefore act in a similar way for parametric locking.



%


\end{document}